\documentclass[aps,prb,twocolumn,letterpaper,superscriptaddress,floatfix]{revtex4-1}

\usepackage{graphicx,amsmath,amsfonts,amssymb,color,ulem}

\def\CaPtFeAs{Ca$_{10}$(Pt$_3$As$_8$)(Fe$_2$As$_2$)$_5$}

\def\CaPtFourFeAs{Ca$_{10}$(Pt$_4$As$_8$)(Fe$_2$As$_2$)$_5$}
\def\BaFeAs{BaFe$_2$As$_2$}

\def\as{$^{75}$As}
\def\be{\begin{equation}}
\def\ee{\end{equation}}

\def\t1{$T_1^{-1}$}

\def\iTone{$T_1^{-1}$}

\def\css{\mathbf{c^{\ast}}}
\begin{document}

\title{Antiferromagnetic order in \CaPtFeAs\ observed by \as\ NMR}

\author{T. Zhou}
\email[]{zhoutong@physics.ucla.edu}
\affiliation{Department of Physics $\&$ Astronomy, UCLA, Los Angeles, CA 90095}
\author{G. Koutroulakis}
\affiliation{Los Alamos National Laboratory, Los Alamos, NM 87545}
\author{J. Lodico}
\affiliation{Department of Physics $\&$ Astronomy, UCLA, Los Angeles, CA 90095}
\author{Ni Ni}
\altaffiliation[current address: ]{Los Alamos National Laboratory, Los Alamos, NM 87545}
\affiliation{Department of Chemistry, Princeton University, Princeton, NJ 08544}
\author{J. D. Thompson}
\affiliation{Los Alamos National Laboratory, Los Alamos, NM 87545}

\author{R. J. Cava}
\affiliation{Department of Chemistry, Princeton University, Princeton, NJ 08544}
\author{S. E. Brown}
\affiliation{Department of Physics $\&$ Astronomy, UCLA, Los Angeles, CA 90095}

\voffset=0.5cm
\begin{abstract}
\as\  nuclear magnetic resonance (NMR) measurements carried out on underdoped, non-superconducting \CaPtFeAs\ reveal physical properties that are similar but not identical to the 122 superconductor parent compounds such as \BaFeAs. Results from the single crystal study indicate a phase transition to an antiferromagnetic (AF) state on cooling through $T\sim100$K, albeit nonuniformly.  Specifically, the NMR lineshape reflects the presence of staggered hyperfine fields on the As sites associated with a striped AF order. The variation of the internal hyperfine field with temperature suggests that the phase transition to the AF state is discontinuous, and therefore likely coincident with the structural transition inferred from transport experiments.
\end{abstract}


\maketitle

The report of superconductivity with $T_c$=26K in the iron-based oxy-pnictide compound LaO$_{1-x}$F$_x$FeAs\cite{Kamihara:2008} initiated exploration of a new family of high-$T_c$ superconductors. Soon after, critical temperatures as high as 55K were found by substituting smaller rare earth elements ({\it e.g.}, Gd \cite{Yang:2008}, Nd \cite{Kito:2008} or Sm \cite{Ren:2008}) for La in the same so-called 1111-structure. As the search for new FeAs-based superconductors continued, the synthesis of the $A$Fe$_2$As$_2$ materials \cite{Rotter:2008,Sasmal:2008,Wu:2008} (forming in the 122-type structure, $A$=Ba,Sr) represented an important step forward for experimental efforts. In these materials, large and good-quality single crystals could be grown by the self-flux method\cite{Ni:2008} and, moreover, the introduction of carriers through various doping strategies could in many cases be well-controlled. The combination of these two factors allowed for systematic studies of the physical properties of these compounds, as well as for the in-depth exploration of hotly debated problems such as the interplay between magnetism and superconductivity \cite{Chen:2009,Pratt:2009}.

Both the parent compounds LaOFeAs and BaFe$_2$As$_2$ exhibit a structural transition, followed by or coincident with striped antiferromagnetic (AF) order. For LaOFeAs, a transition from tetragonal to orthorhombic structure takes place at $T_s=$155K, and an AF transition happens at a lower temperature, $T_N=$142K  \cite{delaCruz:2008}.  In contrast, in \BaFeAs, both transitions take place at the same temperature, $T_s=T_N=$140K \cite{Nakai:2008b}, but $T_N<T_s$ after electron-doping, i.e. replacing Fe for either Co or Ni \cite{Pratt:2009,Lester:2009}. Nevertheless, in both the 1111-type and the 122-type compounds, it has been suggested that the suppression of the AF transition by chemical doping is associated with the formation of the superconducting ground state \cite{Ni:2008,Chu:2009,Ning:2009}.

\begin{figure}[h]
\includegraphics[width=3 in]{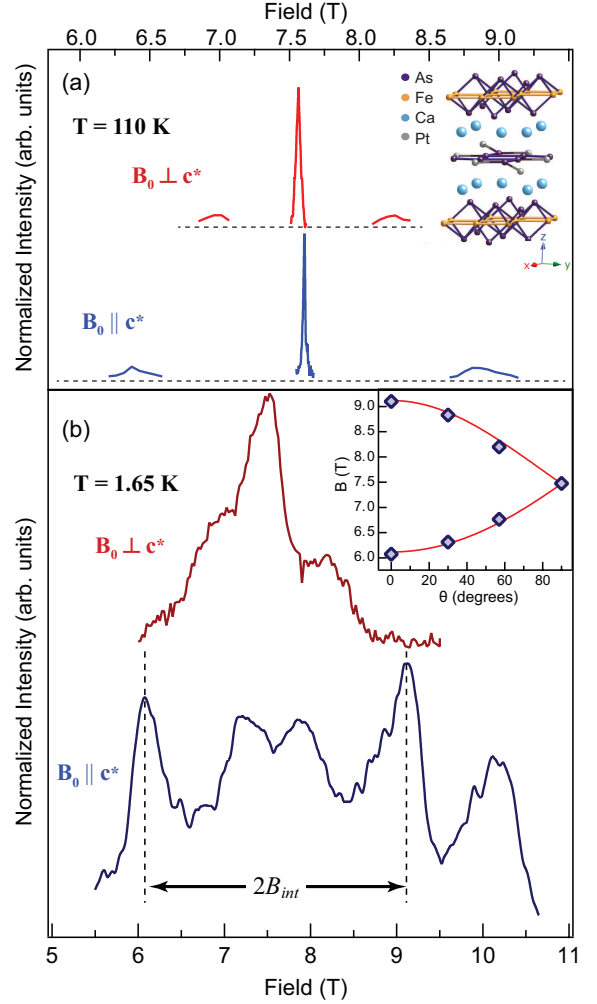}
\caption{\as\ field-swept spectrum for $\mathbf{B_0}(\parallel,\perp) \mathbf{\css}$ and rf carrier frequency $\nu$=55.55MHz: (a) In the paramagnetic phase at $T$=110K. The crystal structure showing the Ca-skutterudite-Ca spacer appears in the inset. (b) In the magnetic phase at $T$=1.65K. The inset shows the angular evolution of the spectrum's magnetically split central transition (blue symbols), and the expected variation for the AF order discussed in the text (red lines).}
\label{fig:spectra}
\end{figure}

Recently, two more complex superconductors in the Ca-Pt-Fe-As chemical system have been found and synthesized, allowing for the study of Fe-based superconductors with weaker interlayer dispersion \cite{Ni:2011,Xiang:2012,Loehnert:2011}. The first, \CaPtFeAs\ (the 10-3-8 phase), has triclinic crystal structure, while the other, \CaPtFourFeAs\ (the 10-4-8 phase), possesses higher, tetragonal symmetry. In both compounds, the skutterudite spacer layer (Pt-As) is sandwiched by Ca ions, as shown in the inset of Fig. \ref{fig:spectra}a, which leads to a significantly greater distance between neighboring FeAs layers than reported previously. The superconducting critical temperature can be tuned by Pt substitution on the Fe site, with its highest value being $T_c=$38K \cite{Kakiya:2011}. In the case of 10-3-8, ($\alpha,\beta,\gamma$)=(94.74,104.34,90.00) \cite{Ni:2011,Nohara:2012}, and (a,c)=(8.759,10.641) with $a=b$. Below, we write the c$^{\ast}$ direction such that $\mathbf{c^{\ast}}\perp(ab)$, and, for the magnetic field $\mathbf{B_0}$ applied along this direction, it is $\mathbf{B_0}\parallel\css\Rightarrow \theta=0$.

A break in slope in the resistivity-temperature curve at $T_s\sim$100K in the undoped or lightly-doped 10-3-8 material indicates a structural transition\cite{Xiang:2012}, which was also verified by polarized-light optical images\cite{Cho:2012}. A superconducting ground state is found at higher carrier concentrations, $x>0.025-0.030$, where $x$ is the fraction of the Fe sites occupied by Pt, and the onset appears coincident with complete suppression of the structural transition. Nevertheless, no accompanying anomaly in the magnetic susceptibility has yet been reported \cite{Ni:2011,Xiang:2012}, casting doubt on the presence of a magnetic ground state. Consequently, it was suggested that the weak interlayer exchange suppresses the long-range magnetic order \cite{Xiang:2012}, even if AF spin fluctuations remain significant for producing superconductivity. Nuclear magnetic resonance (NMR) measurements serve as a local probe of the magnetic state, and, thus, provide an ideal tool for the investigation of the 10-3-8 system, and specifically to determine whether the ground state is antiferromagnetic.

Here we report \as\ NMR experiments on a \CaPtFeAs\ single crystal for a wide range of temperature ($T=$1.65K-115K) and applied magnetic field ($B_0=$5T-11T) values. Most of the results are qualitatively and quantitatively similar to that obtained from the 122 compounds, whereas some differences are observed in the temperature dependence of the spin lattice relaxation rate ($1/T_1$).

\begin{figure}[b]
\includegraphics[width=3 in]{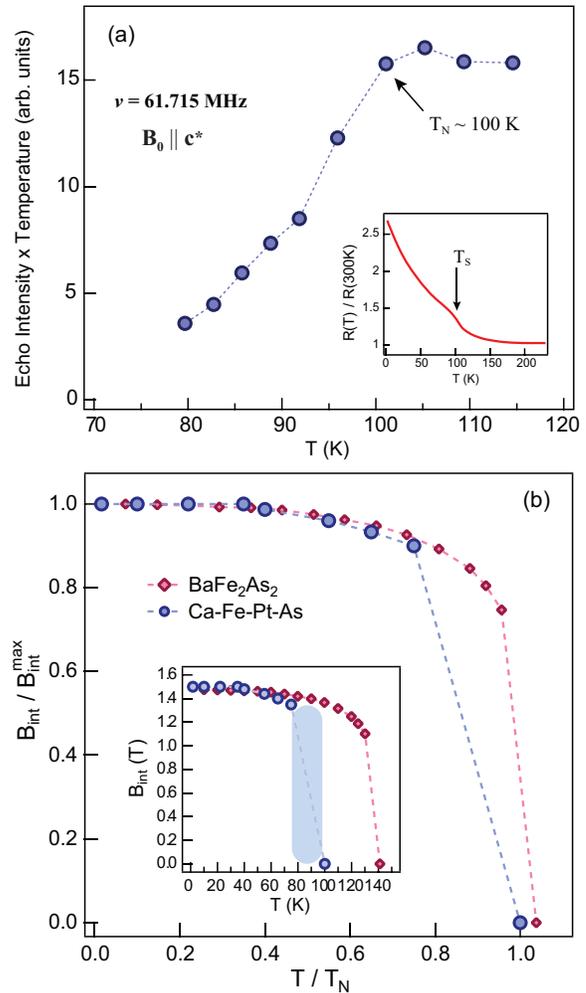}
\caption{(a) Echo intensity vs. $T$, measured under conditions $\mathbf{B_0}=$8.4524T$\css$  and rf carrier $\nu$=61.715MHz. A notable drop in intensity occurs below $T$=100K, which is interpreted as marking the onset of AF order. The inset plots the in-plane resistance $R(T)$, normalized to its room temperature value, for another crystal in the same preparation batch. \newline(b) Internal field measured at the \as\ site for $\theta=0^{\circ}$, normalized to the low-temperature value $B_{int}(T\to0)\simeq$1.5T. The results are contrasted to those obtained for BaFe$_2$As$_2$ \cite{Kitagawa:2008}. Axes are normalized to the internal field for $T\to0$, and N\'eel temperature, respectively. Inset: Unnormalized data. The shaded area highlights the temperature range where a detailed spectrum was impossible to derive due to the spatially inhomogeneous emergence of the AF order within the sample.}
\label{fig:OP}
\end{figure}

The \CaPtFeAs\ single crystal reported on here was prepared and characterized using methods described previously \cite{Ni:2011}. In this case, nominal Pt substitution at the Fe sites leads to  a doping level of $x=0.004\pm0.002$ as determined by wavelength dispersive spectroscopy (WDS) electron probe microanalysis\cite{Cho:2012}. The dimensions of the crystal studied are $\sim$3mm$\times$1mm$\times$0.3mm. The resistance curve shown in the inset of Fig. \ref{fig:OP}a depicts measurements on a crystal from the same batch; transport properties of several additional crystals were examined, and in each case a change in slope in the range 95-105K was observed, suggestive of a phase transition at a temperature $T_s$ that showed variation of approximately $\pm5K$ from one crystal to another.

The coil containing the sample was mounted on a piezo rotator, which allowed for an alignment precision of less than $1^\circ$. The temperature was regulated in a gas-flow $^4$He cryostat with variations limited to $\delta T/T<1\%$. Field-swept NMR spectra were constructed by recording standard Hahn spin echo transients from the \as\ nuclear spins ($I$=3/2). Our expectation was for two very different As environments, since in addition to the FeAs layers, there are distant sites in the Pt-As skutterudite layer (see inset of Fig. \ref{fig:spectra}a). However, only one environment was detected under our experimental conditions. Given the similarity of the results to those obtained for other FeAs-based compounds, we associate our observations with the \as\ nuclear spins nearby the Fe ions. It is likely that the contribution from the skutterudite sites is suppressed due to the weak hyperfine coupling to the Fe spins, which then results in saturation of the associated transitions and very long relaxation times $T_1$. By itself, the weak coupling is not sufficient, since ordinarily the two spin populations will undergo mutual spin-flip transitions as a result of the nuclear dipolar interaction. However, significant hyperfine fields at the sites in the FeAs layers shifts the relative transition frequencies, and could render this mechanism ineffective.

Spectra recorded at $T$=110K and 1.65K are shown in Fig. \ref{fig:spectra}. The spin of the \as\ nucleus is $I$=3/2, and thus its electric quadrupole moment is nonzero and couples to the lattice's electric field gradient (EFG). Hence, the resonance frequencies are given by the eigenvalues of the Hamiltonian
\begin{equation*}
\mathcal{H}=-\gamma \hbar \mathbf{I}\cdot \mathbf{B}_{eff}+\frac{h \nu_Q}{6} \left[ 3 I_z^2 -  I(I+1) + \frac{1}{2}\eta(I_{+}^2 + I_{-}^2) \right].
\end{equation*}
Here, $\gamma=7.292$MHz/T is the \as\ nuclear gyromagnetic ratio, $\nu_Q$ is the quadrupole frequency, $\eta$ the asymmetry parameter, and $\mathbf{B}_{eff}\equiv \mathbf{B_0}+\delta\mathbf{B}$, where $\delta\mathbf{B}$ is the total internal field.  For large magnetic fields ($\gamma \hbar B_{0}\gg h \nu_Q$) aligned with the principle axis $z$ of the EFG, satellite transitions ($I_z=(\pm1/2\leftrightarrow3/2$)) are separated from the central transition ($1/2\leftrightarrow-1/2$) by $\nu_Q$. Here, $z$ is expected to be close to $\css$. For $T=110$K, in the paramagnetic phase, the linewidth (FWHM) of the central transition is $\delta\nu\sim$200kHz (400kHz) for $\mathbf{B_0}\parallel(\perp)\css$, and the satellite transitions are separated from the central by the quadrupole frequency $\nu_Q$=9.0(2)MHz (the NMR parameters are summarized in Table I, along with those of other FeAs parent compounds). Not surprisingly, the linewidths are very broad compared to what is observed for crystals of 1111 or 122 compounds, since due to the low symmetry there are 5 magnetically inequivalent sites just in the FeAs layers, as well as evidence for disorder that will be discussed further below.

\begin{table}[h]
\begin{tabular}{|l|c|c|c|c|c|c|}
\hline
compound & $K_{\parallel}$ (\%) & $K_{\perp}$ (\%) & $\nu_Q$ (MHz)& $T_N$(K)& $B_{int}$ (T) \\
\hline\hline
CaFe$_2$As$_2 \cite{Baek:2009}$ & 0.25 & 0.75& 14 & 167 & 2.6 \\
\hline
BaFe$_2$As$_2$ \cite{Kitagawa:2008} & 0.30 & 0.35& 2.5 & 135 & 1.45 \\
\hline
SrFe$_2$As$_2$ \cite{Kitagawa:2009} & 0.37 & 0.44 & 2.64 & 199 & 2.2\\
\hline
LaOFeAs \cite{Fu:2012,Li:2010} & 0.2 & & 9.2 & 142 &1.6\\
\hline
``10-3-8" & 0.14 & 0.35 & 9.0(2) & 100 & 1.5\\
\hline
\end{tabular}
\label{table:NMRparameters}
\caption{\as\ NMR parameters for familiar parent compounds of Fe-based superconductors, for the purpose of comparison. All values apply to the paramagnetic phase close to the respective $T_N$, with the exception of $B_{int}$, which is referenced to $T\to0$. The \textit{NMR shift} is $K\equiv \mathbf{\delta B} \cdot\mathbf{B}_0/{B_0}^2$. For \CaPtFeAs, the reported value is an average value for the unresolved symmetrically inequivalent sites.  }
\end{table}

At low temperature (Fig. \ref{fig:spectra}b), the spectrum for  \mbox{$\mathbf{B_0}\parallel\css$}, covering the field range 5T$\to$11T, is split into two sets of three maxima\footnote{The sixth, lowest field maximum is not shown due to its overlap with $^{63,65}$Cu contributions arising from the NMR coil}. This results from the hyperfine coupling of the \as\ nuclei to the ordered Fe atomic moments, signaling the presence of long-range AF order. The spectral features are identified and labelled according to nuclear transition, as is the splitting $2B_{int}\simeq$3T arising from the alternating hyperfine field.  The observed peaks are relatively broad, indicating substantial real space variations of the internal field. For $\mathbf{B_0}\perp\css$, the spectrum remains non-split, even though broadened compared to the paramagnetic phase.

The internal field $B_{int}$ corresponds to one half of the peak separation of the corresponding nuclear transitions for $\mathbf{B_0}\parallel\css$. Upon rotation, the spectrum evolves as summarized in the inset of Fig. \ref{fig:spectra}b for the central transition, and any remaining magnetic splitting is unresolved for $\theta=90^{\circ}$. Just as for the 1111\cite{Li:2010} and 122\cite{Zhao:2008,Goldman:2008} compounds, our observations are consistent with striped-phase antiferromagnetism, but with substantially more disorder. That is, the AF order is described by the wave vector $\mathbf{Q}=$(1,0,$n$), with $n$=0,1.

Having established that the underdoped material has an AF ground state, we now consider the identification of the ordering temperature and whether it is distinct from the structural transition at $T_s\sim 100$K. This distinction is difficult to make under the circumstances of broadened spectra, and what we determined to be a spatially inhomogeneous onset of the magnetic order in the crystal employed. The onset of spectral broadening due to static magnetism is evident in Fig. \ref{fig:OP}a, which illustrates the loss of the echo signal intensity originating from the paramagnetic phase. The onset occurs for $T\sim100K$, below which the change is monotonic and continuous. We take the magnetic ordering temperature to coincide with this onset, yielding $T_N=100K$. Note also that the spectral linewidth associated with the intensity shown in Fig. \ref{fig:OP}a remains unchanged even while the intensity is falling so significantly. 

To address the question of whether the transition is continuous or discontinuous, or spatially inhomogeneous, we compare the temperature dependence of the internal field $B_{int}$ to that obtained from the chemically and structurally less complex, clean material \BaFeAs \cite{Kitagawa:2008} in Fig. \ref{fig:OP}b, each normalized to $T_N$. Based on this comparison, the two materials behave analogously, and we therefore conclude that the transition to the low temperature magnetic state is discontinuous for the ideal 10-3-8 phase, and therefore also that it coincides with the structural transition: $T_s=T_N$; in contrast, a significantly smoother falloff of the internal field is seen in Co-doped \BaFeAs, for which $T_N<T_s$  \cite{Ning:2009}. Below we refer to $T_s, T_N$ generally as the transition onset temperature, $T$=100K. Although the order parameter for the structural transition forming below $T_s$ has not been identified, the link with AF striped order is suggestive of a low-temperature triclinic phase (since the high-symmetry phase is already triclinic), but with $a\ne b$.

\begin{figure}[t]
\includegraphics[width=3 in]{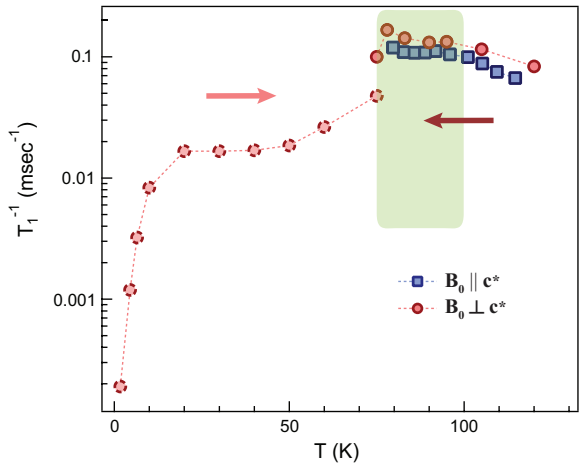}
\caption{\iTone\ vs. $T$ for $\theta=0^{\circ}$ (blue squares), $90^{\circ}$ (red circles). Hashed, lighter (solid, darker) symbols denote data taken going up (down) in  $T$, as indicated by the arrows. The {\it onset} of magnetic order occurs near to 100K. The shaded region illustrates the range 75K$\lesssim T\lesssim$100K where the paramagnetic phase partly survives within the AF phase, due to the sample's chemical inhomogeneity.}
 \label{fig:iTone}
\end{figure}

The variation of the spin lattice relaxation with temperature, shown in Fig. \ref{fig:iTone}, was evaluated by approaching the transition from both higher and lower temperatures. The high-temperature data includes results from the central transition (-1/2$\leftrightarrow$1/2) at $\theta$=0,90$^{\circ}$. From the spectroscopic data, we concluded that the paramagnetic phase continues to exist but in spatially-segregated regions for $T<T_N$. Here, the data for $T>$80K (solid symbols) are associated with the paramagnetic volume fraction. On approaching the transition from lower temperatures from within the magnetic phase (hashed symbols), the measurements were done with $\theta$=90$^{\circ}$, over a temperature range from 1.7K to 75K. For both cases, the values of \iTone\ were defined by comparing the magnetization recovery $M_z(t)$ according to the expectation for magnetic relaxation of the $I=3/2$ central transition,
\be \delta M_z(t)\sim [0.9e^{-6t/T_1}+0.1e^{-t/T_1}]. \label{eq:relaxation}\ee  The data were well-described by the expression Eq. \ref{eq:relaxation} for 10K$\le T\le50$K. For $T<$10K, the reported values were obtained from the same procedure, even though the magnetization recovery was consistent with a distribution of relaxation times. Specifically, a stretched exponential analysis ($\delta M(t)\sim {e^{-(t/T_1)}}^\beta$) results in $\beta=0.5-0.6$. On warming past $T$=50K from the magnetic phase, the data are not well-described using the functional form in Eq. \ref{eq:relaxation}. This discrepancy in the detailed behavior may again be associated with chemical inhomogeneity in the material.

The increase of the relaxation rate upon cooling in the paramagnetic phase is characteristic of AF spin fluctuations and is stronger than for the 122 compounds \cite{Kitagawa:2008,Baek:2009}. Note that the data collected in the paramagnetic phase is associated with a significantly decreasing volume fraction for $T<$100K, the temperature of onset for the spatially inhomogeneous AF order. Once far into the magnetic phase, the relaxation rate first drops by an order of magnitude, but reaches a constant value below 50K. This aspect appears different than the other materials, which exhibit relaxation rates decreasing much more steeply upon cooling. A possible explanation is in fluctuating domains or other dynamical effects, such as those associated with motion of discommensurations. In any case, if the origin is with slow collective dynamics, then the rate will be found dependent on the Larmor frequency.

To conclude, we have performed NMR spectroscopy on a single crystal of non-superconducting \CaPtFeAs. The ground state is antiferromagnetic, with wave vector $\mathbf{Q}=$(1,0,$n$) ($n$=0,1), and forming inhomogeneously below the onset temperature $T_N$=100K. The results are similar in character and amplitude to what is observed in other FeAs compounds that are known to develop striped antiferromagnetism. Variation of the magnetic order parameter with increasing temperature is relatively weak, and reminiscent of what is seen for a first order phase transition. Moreover, since $T_N$ coincides approximately to where anomalies in transport have been associated with a structural transition, we suggest that $T_N\simeq T_s$. Our NMR measurements in the vicinity of $T_N$ are not straightforward to interpret due to the inhomogeneous onset and broadened spectra. A large and temperature-independent relaxation rate of the magnetic phase, extending from $T>$10K is not present in less complex iron arsenides, and possibly originates from the disorder in this system.

The research reported here was supported by the National Science Foundation under grant no. DMR-1105531 (UCLA), and the Air Force Office of Scientific Research Multidisciplinary Research Program for University Research Initiative on superconductivity (Princeton). GK acknowledges support from LANL's Laboratory Directed Research and Development Program and the Seaborg Institute.

Note: The sample quality has been significantly improved recently and further investigation on \CaPtFeAs\ crystals which exhibit clear features in the resistivity, susceptibility and heat capacity is underway.

\begin{thebibliography}{29}
\expandafter\ifx\csname natexlab\endcsname\relax\def\natexlab#1{#1}\fi
\expandafter\ifx\csname bibnamefont\endcsname\relax
  \def\bibnamefont#1{#1}\fi
\expandafter\ifx\csname bibfnamefont\endcsname\relax
  \def\bibfnamefont#1{#1}\fi
\expandafter\ifx\csname citenamefont\endcsname\relax
  \def\citenamefont#1{#1}\fi
\expandafter\ifx\csname url\endcsname\relax
  \def\url#1{\texttt{#1}}\fi
\expandafter\ifx\csname urlprefix\endcsname\relax\def\urlprefix{URL }\fi
\providecommand{\bibinfo}[2]{#2}
\providecommand{\eprint}[2][]{\url{#2}}

\bibitem[{\citenamefont{Kamihara et~al.}(2008)\citenamefont{Kamihara, Watanabe,
  Hirano, and Hosono}}]{Kamihara:2008}
\bibinfo{author}{\bibfnamefont{Y.}~\bibnamefont{Kamihara}},
  \bibinfo{author}{\bibfnamefont{T.}~\bibnamefont{Watanabe}},
  \bibinfo{author}{\bibfnamefont{M.}~\bibnamefont{Hirano}}, \bibnamefont{and}
  \bibinfo{author}{\bibfnamefont{H.}~\bibnamefont{Hosono}},
  \bibinfo{journal}{J. Am. Chem. Soc.} \textbf{\bibinfo{volume}{130}},
  \bibinfo{pages}{3296} (\bibinfo{year}{2008}).

\bibitem[{\citenamefont{Yang et~al.}(2008)\citenamefont{Yang, Li, Lu, Yi, Shen,
  Ren, Che, Dong, Sun, Zhou et~al.}}]{Yang:2008}
\bibinfo{author}{\bibfnamefont{J.}~\bibnamefont{Yang}},
  \bibinfo{author}{\bibfnamefont{Z.-C.} \bibnamefont{Li}},
  \bibinfo{author}{\bibfnamefont{W.}~\bibnamefont{Lu}},
  \bibinfo{author}{\bibfnamefont{W.}~\bibnamefont{Yi}},
  \bibinfo{author}{\bibfnamefont{X.-L.} \bibnamefont{Shen}},
  \bibinfo{author}{\bibfnamefont{Z.-A.} \bibnamefont{Ren}},
  \bibinfo{author}{\bibfnamefont{G.-C.} \bibnamefont{Che}},
  \bibinfo{author}{\bibfnamefont{X.-L.} \bibnamefont{Dong}},
  \bibinfo{author}{\bibfnamefont{L.-L.} \bibnamefont{Sun}},
  \bibinfo{author}{\bibfnamefont{F.}~\bibnamefont{Zhou}}, \bibnamefont{et~al.},
  \bibinfo{journal}{Supercond. Sci. Tech.} \textbf{\bibinfo{volume}{21}},
  \bibinfo{pages}{082001} (\bibinfo{year}{2008}).

\bibitem[{\citenamefont{Kito et~al.}(2008)\citenamefont{Kito, Eisaki, and
  Iyo}}]{Kito:2008}
\bibinfo{author}{\bibfnamefont{H.}~\bibnamefont{Kito}},
  \bibinfo{author}{\bibfnamefont{H.}~\bibnamefont{Eisaki}}, \bibnamefont{and}
  \bibinfo{author}{\bibfnamefont{A.}~\bibnamefont{Iyo}}, \bibinfo{journal}{J.
  Phys. Soc. Japan} \textbf{\bibinfo{volume}{77}}, \bibinfo{pages}{063707}
  (\bibinfo{year}{2008}).

\bibitem[{\citenamefont{Zhi-An et~al.}(2008)\citenamefont{Zhi-An, Wei, Jie,
  Wei, Xiao-Li, Zheng-Cai, Guang-Can, Xiao-Li, Li-Ling, Fang
  et~al.}}]{Ren:2008}
\bibinfo{author}{\bibfnamefont{R.}~\bibnamefont{Zhi-An}},
  \bibinfo{author}{\bibfnamefont{L.}~\bibnamefont{Wei}},
  \bibinfo{author}{\bibfnamefont{Y.}~\bibnamefont{Jie}},
  \bibinfo{author}{\bibfnamefont{Y.}~\bibnamefont{Wei}},
  \bibinfo{author}{\bibfnamefont{S.}~\bibnamefont{Xiao-Li}},
  \bibinfo{author}{\bibnamefont{Zheng-Cai}},
  \bibinfo{author}{\bibfnamefont{C.}~\bibnamefont{Guang-Can}},
  \bibinfo{author}{\bibfnamefont{D.}~\bibnamefont{Xiao-Li}},
  \bibinfo{author}{\bibfnamefont{S.}~\bibnamefont{Li-Ling}},
  \bibinfo{author}{\bibfnamefont{Z.}~\bibnamefont{Fang}}, \bibnamefont{et~al.},
  \bibinfo{journal}{Chin. Phys. Lett.} \textbf{\bibinfo{volume}{25}},
  \bibinfo{pages}{2215} (\bibinfo{year}{2008}).

\bibitem[{\citenamefont{Rotter et~al.}(2008)\citenamefont{Rotter, Tegel,
  Johrendt, Schellenberg, Hermes, and P\"ottgen}}]{Rotter:2008}
\bibinfo{author}{\bibfnamefont{M.}~\bibnamefont{Rotter}},
  \bibinfo{author}{\bibfnamefont{M.}~\bibnamefont{Tegel}},
  \bibinfo{author}{\bibfnamefont{D.}~\bibnamefont{Johrendt}},
  \bibinfo{author}{\bibfnamefont{I.}~\bibnamefont{Schellenberg}},
  \bibinfo{author}{\bibfnamefont{W.}~\bibnamefont{Hermes}}, \bibnamefont{and}
  \bibinfo{author}{\bibfnamefont{R.}~\bibnamefont{P\"ottgen}},
  \bibinfo{journal}{Phys. Rev. B} \textbf{\bibinfo{volume}{78}},
  \bibinfo{pages}{020503} (\bibinfo{year}{2008}).

\bibitem[{\citenamefont{Sasmal et~al.}(2008)\citenamefont{Sasmal, Lv, Lorenz,
  Guloy, Chen, Xue, and Chu}}]{Sasmal:2008}
\bibinfo{author}{\bibfnamefont{K.}~\bibnamefont{Sasmal}},
  \bibinfo{author}{\bibfnamefont{B.}~\bibnamefont{Lv}},
  \bibinfo{author}{\bibfnamefont{B.}~\bibnamefont{Lorenz}},
  \bibinfo{author}{\bibfnamefont{A.~M.} \bibnamefont{Guloy}},
  \bibinfo{author}{\bibfnamefont{F.}~\bibnamefont{Chen}},
  \bibinfo{author}{\bibfnamefont{Y.-Y.} \bibnamefont{Xue}}, \bibnamefont{and}
  \bibinfo{author}{\bibfnamefont{C.-W.} \bibnamefont{Chu}},
  \bibinfo{journal}{Phys. Rev. Lett.} \textbf{\bibinfo{volume}{101}},
  \bibinfo{pages}{107007} (\bibinfo{year}{2008}).

\bibitem[{\citenamefont{Wu et~al.}(2008)\citenamefont{Wu, Liu, Chen, Yan, Wu,
  Xie, Ying, Wang, Fang, and Chen}}]{Wu:2008}
\bibinfo{author}{\bibfnamefont{G.}~\bibnamefont{Wu}},
  \bibinfo{author}{\bibfnamefont{R.~H.} \bibnamefont{Liu}},
  \bibinfo{author}{\bibfnamefont{H.}~\bibnamefont{Chen}},
  \bibinfo{author}{\bibfnamefont{Y.~J.} \bibnamefont{Yan}},
  \bibinfo{author}{\bibfnamefont{T.}~\bibnamefont{Wu}},
  \bibinfo{author}{\bibfnamefont{Y.~L.} \bibnamefont{Xie}},
  \bibinfo{author}{\bibfnamefont{J.~J.} \bibnamefont{Ying}},
  \bibinfo{author}{\bibfnamefont{X.~F.} \bibnamefont{Wang}},
  \bibinfo{author}{\bibfnamefont{D.~F.} \bibnamefont{Fang}}, \bibnamefont{and}
  \bibinfo{author}{\bibfnamefont{X.~H.} \bibnamefont{Chen}},
  \bibinfo{journal}{Europhys. Lett.} \textbf{\bibinfo{volume}{84}},
  \bibinfo{pages}{27010} (\bibinfo{year}{2008}).

\bibitem[{\citenamefont{Ni et~al.}(2008)\citenamefont{Ni, Tillman, Yan,
  Kracher, Hannahs, Bud'ko, and Canfield}}]{Ni:2008}
\bibinfo{author}{\bibfnamefont{N.}~\bibnamefont{Ni}},
  \bibinfo{author}{\bibfnamefont{M.~E.} \bibnamefont{Tillman}},
  \bibinfo{author}{\bibfnamefont{J.-Q.} \bibnamefont{Yan}},
  \bibinfo{author}{\bibfnamefont{A.}~\bibnamefont{Kracher}},
  \bibinfo{author}{\bibfnamefont{S.~T.} \bibnamefont{Hannahs}},
  \bibinfo{author}{\bibfnamefont{S.~L.} \bibnamefont{Bud'ko}},
  \bibnamefont{and} \bibinfo{author}{\bibfnamefont{P.~C.}
  \bibnamefont{Canfield}}, \bibinfo{journal}{Phys. Rev. B}
  \textbf{\bibinfo{volume}{78}}, \bibinfo{pages}{214515}
  (\bibinfo{year}{2008}).

\bibitem[{\citenamefont{Chen et~al.}(2009)\citenamefont{Chen, Ren, Qiu, Bao,
  Liu, Wu, Wu, Xie, Wang, Huang et~al.}}]{Chen:2009}
\bibinfo{author}{\bibfnamefont{H.}~\bibnamefont{Chen}},
  \bibinfo{author}{\bibfnamefont{Y.}~\bibnamefont{Ren}},
  \bibinfo{author}{\bibfnamefont{Y.}~\bibnamefont{Qiu}},
  \bibinfo{author}{\bibfnamefont{W.}~\bibnamefont{Bao}},
  \bibinfo{author}{\bibfnamefont{R.~H.} \bibnamefont{Liu}},
  \bibinfo{author}{\bibfnamefont{G.}~\bibnamefont{Wu}},
  \bibinfo{author}{\bibfnamefont{T.}~\bibnamefont{Wu}},
  \bibinfo{author}{\bibfnamefont{Y.~L.} \bibnamefont{Xie}},
  \bibinfo{author}{\bibfnamefont{X.~F.} \bibnamefont{Wang}},
  \bibinfo{author}{\bibfnamefont{Q.}~\bibnamefont{Huang}},
  \bibnamefont{et~al.}, \bibinfo{journal}{Europhys. Lett.}
  \textbf{\bibinfo{volume}{85}}, \bibinfo{pages}{17006} (\bibinfo{year}{2009}).

\bibitem[{\citenamefont{Pratt et~al.}(2009)\citenamefont{Pratt, Tian, Kreyssig,
  Zarestky, Nandi, Ni, Bud'ko, Canfield, Goldman, and McQueeney}}]{Pratt:2009}
\bibinfo{author}{\bibfnamefont{D.~K.} \bibnamefont{Pratt}},
  \bibinfo{author}{\bibfnamefont{W.}~\bibnamefont{Tian}},
  \bibinfo{author}{\bibfnamefont{A.}~\bibnamefont{Kreyssig}},
  \bibinfo{author}{\bibfnamefont{J.~L.} \bibnamefont{Zarestky}},
  \bibinfo{author}{\bibfnamefont{S.}~\bibnamefont{Nandi}},
  \bibinfo{author}{\bibfnamefont{N.}~\bibnamefont{Ni}},
  \bibinfo{author}{\bibfnamefont{S.~L.} \bibnamefont{Bud'ko}},
  \bibinfo{author}{\bibfnamefont{P.~C.} \bibnamefont{Canfield}},
  \bibinfo{author}{\bibfnamefont{A.~I.} \bibnamefont{Goldman}},
  \bibnamefont{and} \bibinfo{author}{\bibfnamefont{R.~J.}
  \bibnamefont{McQueeney}}, \bibinfo{journal}{Phys. Rev. Lett.}
  \textbf{\bibinfo{volume}{103}}, \bibinfo{pages}{087001}
  (\bibinfo{year}{2009}).

\bibitem[{\citenamefont{de~la Cruz et~al.}(2008)\citenamefont{de~la Cruz,
  Huang, Lynn, Li, II, Zarestky, Mook, Chen, Luo, Wang et~al.}}]{delaCruz:2008}
\bibinfo{author}{\bibfnamefont{C.}~\bibnamefont{de~la Cruz}},
  \bibinfo{author}{\bibfnamefont{Q.}~\bibnamefont{Huang}},
  \bibinfo{author}{\bibfnamefont{J.~W.} \bibnamefont{Lynn}},
  \bibinfo{author}{\bibfnamefont{J.}~\bibnamefont{Li}},
  \bibinfo{author}{\bibfnamefont{W.~R.} \bibnamefont{II}},
  \bibinfo{author}{\bibfnamefont{J.~L.} \bibnamefont{Zarestky}},
  \bibinfo{author}{\bibfnamefont{H.~A.} \bibnamefont{Mook}},
  \bibinfo{author}{\bibfnamefont{G.~F.} \bibnamefont{Chen}},
  \bibinfo{author}{\bibfnamefont{J.~L.} \bibnamefont{Luo}},
  \bibinfo{author}{\bibfnamefont{N.~L.} \bibnamefont{Wang}},
  \bibnamefont{et~al.}, \bibinfo{journal}{Nature}
  \textbf{\bibinfo{volume}{453}}, \bibinfo{pages}{899} (\bibinfo{year}{2008}).

\bibitem[{\citenamefont{Nakai et~al.}(2008)\citenamefont{Nakai, Ishida,
  Kamihara, Hirano, and Hosono}}]{Nakai:2008b}
\bibinfo{author}{\bibfnamefont{Y.}~\bibnamefont{Nakai}},
  \bibinfo{author}{\bibfnamefont{K.}~\bibnamefont{Ishida}},
  \bibinfo{author}{\bibfnamefont{Y.}~\bibnamefont{Kamihara}},
  \bibinfo{author}{\bibfnamefont{M.}~\bibnamefont{Hirano}}, \bibnamefont{and}
  \bibinfo{author}{\bibfnamefont{H.}~\bibnamefont{Hosono}},
  \bibinfo{journal}{J. Phys. Soc. Japan} \textbf{\bibinfo{volume}{77}},
  \bibinfo{pages}{073701} (\bibinfo{year}{2008}).

\bibitem[{\citenamefont{Lester et~al.}(2009)\citenamefont{Lester, Chu,
  Analytis, Capelli, Erickson, Condron, Toney, Fisher, and
  Hayden}}]{Lester:2009}
\bibinfo{author}{\bibfnamefont{C.}~\bibnamefont{Lester}},
  \bibinfo{author}{\bibfnamefont{J.-H.} \bibnamefont{Chu}},
  \bibinfo{author}{\bibfnamefont{J.~G.} \bibnamefont{Analytis}},
  \bibinfo{author}{\bibfnamefont{S.~C.} \bibnamefont{Capelli}},
  \bibinfo{author}{\bibfnamefont{A.~S.} \bibnamefont{Erickson}},
  \bibinfo{author}{\bibfnamefont{C.~L.} \bibnamefont{Condron}},
  \bibinfo{author}{\bibfnamefont{M.~F.} \bibnamefont{Toney}},
  \bibinfo{author}{\bibfnamefont{I.~R.} \bibnamefont{Fisher}},
  \bibnamefont{and} \bibinfo{author}{\bibfnamefont{S.~M.}
  \bibnamefont{Hayden}}, \bibinfo{journal}{Phys. Rev. B}
  \textbf{\bibinfo{volume}{79}}, \bibinfo{pages}{144523}
  (\bibinfo{year}{2009}).

\bibitem[{\citenamefont{Chu et~al.}(2009)\citenamefont{Chu, Analytis,
  Kucharczyk, and Fisher}}]{Chu:2009}
\bibinfo{author}{\bibfnamefont{J.-H.} \bibnamefont{Chu}},
  \bibinfo{author}{\bibfnamefont{J.~G.} \bibnamefont{Analytis}},
  \bibinfo{author}{\bibfnamefont{C.}~\bibnamefont{Kucharczyk}},
  \bibnamefont{and} \bibinfo{author}{\bibfnamefont{I.~R.}
  \bibnamefont{Fisher}}, \bibinfo{journal}{Phys. Rev. B}
  \textbf{\bibinfo{volume}{79}}, \bibinfo{pages}{014506}
  (\bibinfo{year}{2009}).

\bibitem[{\citenamefont{Ning et~al.}(2009)\citenamefont{Ning, Ahilan, Imai,
  Sefat, Jin, McGuire, Sales, and Mandrus}}]{Ning:2009}
\bibinfo{author}{\bibfnamefont{F.}~\bibnamefont{Ning}},
  \bibinfo{author}{\bibfnamefont{K.}~\bibnamefont{Ahilan}},
  \bibinfo{author}{\bibfnamefont{T.}~\bibnamefont{Imai}},
  \bibinfo{author}{\bibfnamefont{A.~S.} \bibnamefont{Sefat}},
  \bibinfo{author}{\bibfnamefont{R.}~\bibnamefont{Jin}},
  \bibinfo{author}{\bibfnamefont{M.~A.} \bibnamefont{McGuire}},
  \bibinfo{author}{\bibfnamefont{B.~C.} \bibnamefont{Sales}}, \bibnamefont{and}
  \bibinfo{author}{\bibfnamefont{D.}~\bibnamefont{Mandrus}},
  \bibinfo{journal}{J. Phys. Soc. Japan} \textbf{\bibinfo{volume}{78}},
  \bibinfo{pages}{013711} (\bibinfo{year}{2009}).

\bibitem[{\citenamefont{Ni et~al.}(2011)\citenamefont{Ni, Allred, Chan, and
  Cava}}]{Ni:2011}
\bibinfo{author}{\bibfnamefont{N.}~\bibnamefont{Ni}},
  \bibinfo{author}{\bibfnamefont{J.~M.} \bibnamefont{Allred}},
  \bibinfo{author}{\bibfnamefont{B.~C.} \bibnamefont{Chan}}, \bibnamefont{and}
  \bibinfo{author}{\bibfnamefont{R.~J.} \bibnamefont{Cava}},
  \bibinfo{journal}{Proc. Nat. Acad. Sci.} \textbf{\bibinfo{volume}{108}},
  \bibinfo{pages}{E1019} (\bibinfo{year}{2011}).

\bibitem[{\citenamefont{Xiang et~al.}(2012)\citenamefont{Xiang, Luo, Ying,
  Wang, Yan, Wang, Cheng, Ye, and Chen}}]{Xiang:2012}
\bibinfo{author}{\bibfnamefont{Z.~J.} \bibnamefont{Xiang}},
  \bibinfo{author}{\bibfnamefont{X.~G.} \bibnamefont{Luo}},
  \bibinfo{author}{\bibfnamefont{J.~J.} \bibnamefont{Ying}},
  \bibinfo{author}{\bibfnamefont{X.~F.} \bibnamefont{Wang}},
  \bibinfo{author}{\bibfnamefont{Y.~J.} \bibnamefont{Yan}},
  \bibinfo{author}{\bibfnamefont{A.~F.} \bibnamefont{Wang}},
  \bibinfo{author}{\bibfnamefont{P.}~\bibnamefont{Cheng}},
  \bibinfo{author}{\bibfnamefont{G.~J.} \bibnamefont{Ye}}, \bibnamefont{and}
  \bibinfo{author}{\bibfnamefont{X.~H.} \bibnamefont{Chen}},
  \bibinfo{journal}{Phys. Rev. B} \textbf{\bibinfo{volume}{85}},
  \bibinfo{pages}{224527} (\bibinfo{year}{2012}).

\bibitem[{\citenamefont{L\"ohnert et~al.}(2011)\citenamefont{L\"ohnert,
  St\"urzer, Tegel, Frankovsky, Friederichs, and Johrendt}}]{Loehnert:2011}
\bibinfo{author}{\bibfnamefont{C.}~\bibnamefont{L\"ohnert}},
  \bibinfo{author}{\bibfnamefont{T.}~\bibnamefont{St\"urzer}},
  \bibinfo{author}{\bibfnamefont{M.}~\bibnamefont{Tegel}},
  \bibinfo{author}{\bibfnamefont{R.}~\bibnamefont{Frankovsky}},
  \bibinfo{author}{\bibfnamefont{G.}~\bibnamefont{Friederichs}},
  \bibnamefont{and} \bibinfo{author}{\bibfnamefont{D.}~\bibnamefont{Johrendt}},
  \bibinfo{journal}{Angewandte Chemie Int. Ed.} \textbf{\bibinfo{volume}{50}},
  \bibinfo{pages}{9195} (\bibinfo{year}{2011}).

\bibitem[{\citenamefont{Kakiya et~al.}(2011)\citenamefont{Kakiya, Kudo,
  Nishikubo, Oku, Nishibori, Sawa, Yamamoto, Nozaka, and Nohara}}]{Kakiya:2011}
\bibinfo{author}{\bibfnamefont{S.}~\bibnamefont{Kakiya}},
  \bibinfo{author}{\bibfnamefont{K.}~\bibnamefont{Kudo}},
  \bibinfo{author}{\bibfnamefont{Y.}~\bibnamefont{Nishikubo}},
  \bibinfo{author}{\bibfnamefont{K.}~\bibnamefont{Oku}},
  \bibinfo{author}{\bibfnamefont{E.}~\bibnamefont{Nishibori}},
  \bibinfo{author}{\bibfnamefont{H.}~\bibnamefont{Sawa}},
  \bibinfo{author}{\bibfnamefont{T.}~\bibnamefont{Yamamoto}},
  \bibinfo{author}{\bibfnamefont{T.}~\bibnamefont{Nozaka}}, \bibnamefont{and}
  \bibinfo{author}{\bibfnamefont{M.}~\bibnamefont{Nohara}},
  \bibinfo{journal}{J. Phys. Soc. Japan} \textbf{\bibinfo{volume}{80}},
  \bibinfo{pages}{093704} (\bibinfo{year}{2011}).

\bibitem[{\citenamefont{Nohara et~al.}(2012)\citenamefont{Nohara, Kakiya, Kudo,
  Oshiro, Araki, Kobayashi, Oku, Nishibori, and Sawa}}]{Nohara:2012}
\bibinfo{author}{\bibfnamefont{M.}~\bibnamefont{Nohara}},
  \bibinfo{author}{\bibfnamefont{S.}~\bibnamefont{Kakiya}},
  \bibinfo{author}{\bibfnamefont{K.}~\bibnamefont{Kudo}},
  \bibinfo{author}{\bibfnamefont{Y.}~\bibnamefont{Oshiro}},
  \bibinfo{author}{\bibfnamefont{S.}~\bibnamefont{Araki}},
  \bibinfo{author}{\bibfnamefont{T.~C.} \bibnamefont{Kobayashi}},
  \bibinfo{author}{\bibfnamefont{K.}~\bibnamefont{Oku}},
  \bibinfo{author}{\bibfnamefont{E.}~\bibnamefont{Nishibori}},
  \bibnamefont{and} \bibinfo{author}{\bibfnamefont{H.}~\bibnamefont{Sawa}},
  \bibinfo{journal}{Solid State Commun.} \textbf{\bibinfo{volume}{152}},
  \bibinfo{pages}{635} (\bibinfo{year}{2012}).

\bibitem[{\citenamefont{Cho et~al.}(2012)\citenamefont{Cho, Tanatar, Kim,
  Straszheim, Ni, Cava, and Prozorov}}]{Cho:2012}
\bibinfo{author}{\bibfnamefont{K.}~\bibnamefont{Cho}},
  \bibinfo{author}{\bibfnamefont{M.~A.} \bibnamefont{Tanatar}},
  \bibinfo{author}{\bibfnamefont{H.}~\bibnamefont{Kim}},
  \bibinfo{author}{\bibfnamefont{W.~E.} \bibnamefont{Straszheim}},
  \bibinfo{author}{\bibfnamefont{N.}~\bibnamefont{Ni}},
  \bibinfo{author}{\bibfnamefont{R.~J.} \bibnamefont{Cava}}, \bibnamefont{and}
  \bibinfo{author}{\bibfnamefont{R.}~\bibnamefont{Prozorov}},
  \bibinfo{journal}{Phys. Rev. B} \textbf{\bibinfo{volume}{85}},
  \bibinfo{pages}{020504} (\bibinfo{year}{2012}).

\bibitem[{\citenamefont{Baek et~al.}(2009)\citenamefont{Baek, Curro, Klimczuk,
  Bauer, Ronning, and Thompson}}]{Baek:2009}
\bibinfo{author}{\bibfnamefont{S.-H.} \bibnamefont{Baek}},
  \bibinfo{author}{\bibfnamefont{N.~J.} \bibnamefont{Curro}},
  \bibinfo{author}{\bibfnamefont{T.}~\bibnamefont{Klimczuk}},
  \bibinfo{author}{\bibfnamefont{E.~D.} \bibnamefont{Bauer}},
  \bibinfo{author}{\bibfnamefont{F.}~\bibnamefont{Ronning}}, \bibnamefont{and}
  \bibinfo{author}{\bibfnamefont{J.~D.} \bibnamefont{Thompson}},
  \bibinfo{journal}{Phys. Rev. B} \textbf{\bibinfo{volume}{79}},
  \bibinfo{pages}{052504} (\bibinfo{year}{2009}).

\bibitem[{\citenamefont{Kitagawa et~al.}(2008)\citenamefont{Kitagawa, Katayama,
  Ohgushi, Yoshida, and Takigawa}}]{Kitagawa:2008}
\bibinfo{author}{\bibfnamefont{K.}~\bibnamefont{Kitagawa}},
  \bibinfo{author}{\bibfnamefont{N.}~\bibnamefont{Katayama}},
  \bibinfo{author}{\bibfnamefont{K.}~\bibnamefont{Ohgushi}},
  \bibinfo{author}{\bibfnamefont{M.}~\bibnamefont{Yoshida}}, \bibnamefont{and}
  \bibinfo{author}{\bibfnamefont{M.}~\bibnamefont{Takigawa}},
  \bibinfo{journal}{J. Phys. Soc. Japan} \textbf{\bibinfo{volume}{77}},
  \bibinfo{pages}{114709} (\bibinfo{year}{2008}).

\bibitem[{\citenamefont{Kitagawa et~al.}(2009)\citenamefont{Kitagawa, Katayama,
  Ohgushi, and Takigawa}}]{Kitagawa:2009}
\bibinfo{author}{\bibfnamefont{K.}~\bibnamefont{Kitagawa}},
  \bibinfo{author}{\bibfnamefont{N.}~\bibnamefont{Katayama}},
  \bibinfo{author}{\bibfnamefont{K.}~\bibnamefont{Ohgushi}}, \bibnamefont{and}
  \bibinfo{author}{\bibfnamefont{M.}~\bibnamefont{Takigawa}},
  \bibinfo{journal}{J. Phys. Soc. Japan} \textbf{\bibinfo{volume}{78}},
  \bibinfo{pages}{063706} (\bibinfo{year}{2009}).

\bibitem[{\citenamefont{Fu et~al.}(2009)\citenamefont{Fu, Torchetti, Imai,
  Ning, Yan, and Sefat}}]{Fu:2012}
\bibinfo{author}{\bibfnamefont{M.}~\bibnamefont{Fu}},
  \bibinfo{author}{\bibfnamefont{D.~A.} \bibnamefont{Torchetti}},
  \bibinfo{author}{\bibfnamefont{T.}~\bibnamefont{Imai}},
  \bibinfo{author}{\bibfnamefont{F.~L.} \bibnamefont{Ning}},
  \bibinfo{author}{\bibfnamefont{J.-Q.} \bibnamefont{Yan}}, \bibnamefont{and}
  \bibinfo{author}{\bibfnamefont{A.~S.} \bibnamefont{Sefat}},
  \bibinfo{journal}{arXiv:1208.5652v1}  (\bibinfo{year}{2009}).

\bibitem[{\citenamefont{Li et~al.}(2010)\citenamefont{Li, Tian, Yan, Zarestky,
  McCallum, Lograsso, and Vaknin}}]{Li:2010}
\bibinfo{author}{\bibfnamefont{H.-F.} \bibnamefont{Li}},
  \bibinfo{author}{\bibfnamefont{W.}~\bibnamefont{Tian}},
  \bibinfo{author}{\bibfnamefont{J.-Q.} \bibnamefont{Yan}},
  \bibinfo{author}{\bibfnamefont{J.~L.} \bibnamefont{Zarestky}},
  \bibinfo{author}{\bibfnamefont{R.~W.} \bibnamefont{McCallum}},
  \bibinfo{author}{\bibfnamefont{T.~A.} \bibnamefont{Lograsso}},
  \bibnamefont{and} \bibinfo{author}{\bibfnamefont{D.}~\bibnamefont{Vaknin}},
  \bibinfo{journal}{Phys. Rev. B} \textbf{\bibinfo{volume}{82}},
  \bibinfo{pages}{064409} (\bibinfo{year}{2010}).

\bibitem[{Note1()}]{Note1}
\bibinfo{note}{The sixth, lowest field maximum is not shown due to its
  overlap with $^{63,65}$Cu contributions arising from the NMR coil}.

\bibitem[{\citenamefont{Zhao et~al.}(2008)\citenamefont{Zhao, Ratcliff, Lynn,
  Chen, Luo, Wang, Hu, and Dai}}]{Zhao:2008}
\bibinfo{author}{\bibfnamefont{J.}~\bibnamefont{Zhao}},
  \bibinfo{author}{\bibfnamefont{W.}~\bibnamefont{Ratcliff}},
  \bibinfo{author}{\bibfnamefont{J.~W.} \bibnamefont{Lynn}},
  \bibinfo{author}{\bibfnamefont{G.~F.} \bibnamefont{Chen}},
  \bibinfo{author}{\bibfnamefont{J.~L.} \bibnamefont{Luo}},
  \bibinfo{author}{\bibfnamefont{N.~L.} \bibnamefont{Wang}},
  \bibinfo{author}{\bibfnamefont{J.}~\bibnamefont{Hu}}, \bibnamefont{and}
  \bibinfo{author}{\bibfnamefont{P.}~\bibnamefont{Dai}},
  \bibinfo{journal}{Phys. Rev. B} \textbf{\bibinfo{volume}{78}},
  \bibinfo{pages}{140504} (\bibinfo{year}{2008}).

\bibitem[{\citenamefont{Goldman et~al.}(2008)\citenamefont{Goldman, Argyriou,
  Ouladdiaf, Chatterji, Kreyssig, Nandi, Ni, Bud'ko, Canfield, and
  McQueeney}}]{Goldman:2008}
\bibinfo{author}{\bibfnamefont{A.~I.} \bibnamefont{Goldman}},
  \bibinfo{author}{\bibfnamefont{D.~N.} \bibnamefont{Argyriou}},
  \bibinfo{author}{\bibfnamefont{B.}~\bibnamefont{Ouladdiaf}},
  \bibinfo{author}{\bibfnamefont{T.}~\bibnamefont{Chatterji}},
  \bibinfo{author}{\bibfnamefont{A.}~\bibnamefont{Kreyssig}},
  \bibinfo{author}{\bibfnamefont{S.}~\bibnamefont{Nandi}},
  \bibinfo{author}{\bibfnamefont{N.}~\bibnamefont{Ni}},
  \bibinfo{author}{\bibfnamefont{S.~L.} \bibnamefont{Bud'ko}},
  \bibinfo{author}{\bibfnamefont{P.~C.} \bibnamefont{Canfield}},
  \bibnamefont{and} \bibinfo{author}{\bibfnamefont{R.~J.}
  \bibnamefont{McQueeney}}, \bibinfo{journal}{Phys. Rev. B}
  \textbf{\bibinfo{volume}{78}}, \bibinfo{pages}{100506}
  (\bibinfo{year}{2008}).

\end{thebibliography}

\end{document}